\documentclass[12pt,a4paper]{{amsart}}
\usepackage{verbatim} 
\usepackage{amssymb,amsfonts,latexsym,amscd}
\newcommand{\di}{\,\mathrm{d}}
\newcommand{\R}{\mathbb{R}}
\newcommand{\Hn}{\mathbb{H}^n}
\newcommand{\tHn}{\mathbb{\tilde H}^n}
\newcommand{\Grad}{\textmd{Grad }}
\newcommand{\dvg}{\textmd{div }}
\newcommand{\tmax}{{\tilde \theta}}

\newtheorem{Theorem}{Theorem}[section]
\newtheorem{Lemma}{Lemma}[section]
\newtheorem{Fact}{Fact}[section]

\theoremstyle{remark}
\newtheorem*{Remark}{Remark}

\author{Rafael D. Benguria \and Helmut Linde} %
\email{RBenguri@fis.puc.cl, Helmut.Linde@gmx.de}
\title[A second eigenvalue bound]
{A second eigenvalue bound for the Dirichlet Laplacian in hyperbolic space}
\address{Department of Physics, Pontific\'\i a Universidad Cat\'olica de Chile Casilla 306, Correo 22
Santiago, Chile.}
\thanks{R.B. was supported by FONDECYT project \# 102-0844. H.L. gratefully acknowledges financial support from CONICYT}

\begin{document}

\begin{abstract}
Let $\Omega$ be some domain in the hyperbolic space $\Hn$ (with $n\ge 2$) and $S_1$ the geodesic ball that has the same
first Dirichlet eigenvalue as $\Omega$. We prove the Payne-P\'olya-Weinberger conjecture for $\Hn$, i.e., that the
second Dirichlet eigenvalue on $\Omega$ is smaller or equal than the second Dirichlet eigenvalue on $S_1$.

We also prove that the ratio of the first two eigenvalues on geodesic balls is a decreasing function of the radius.
\end{abstract}
\maketitle

\section{Introduction} \label{SectionIntroduction}

In the proof of the Payne-P\'olya-Weinberger (PPW) conjecture \cite{AB92a}, Ashbaugh and one of us showed that the
ratio of the first two Dirichlet eigenvalues of the Laplacian operator on a bounded domain in Euclidean space is
maximized when the domain is a circle. This result was later generalized to domains that are contained in a hemisphere
of $\mathbb S^n$ \cite{AB00} and, more recently, to Schr\"odinger operators in the Euclidean space and to the Dirichlet
Laplacian in Gaussian space \cite{H,BL06}. In the work at hand we prove an analogous result for the hyperbolic space
that has already been conjectured in \cite{AB00}:
\begin{Theorem}[The PPW inequality for $\Hn$]  \label{TheoremPPW}
Let $\Omega \subset \Hn$ be an open bounded domain in the hyperbolic space of constant negative curvature
$\kappa\equiv-\rho^{-2}$ and call $\lambda_i(\Omega,\rho)$ the $i$--th Dirichlet eigenvalue on $\Omega$. If $S_1
\subset\Hn$ is a geodesic ball such that $\lambda_1(\Omega,\rho) = \lambda_1(S_1,\rho)$ then
\begin{equation}\label{EqMainResult}
\lambda_2(\Omega,\rho) \le \lambda_2(S_1,\rho)
\end{equation}
with equality if and only if $\Omega$ is a geodesic ball.
\end{Theorem}
For the precise definitions of $\Hn$ and the Laplacian operator in it, see the following section.

We remark that the inequality (\ref{EqMainResult}) has not the form of the original PPW estimate
\begin{equation}\label{EqOriginalPPW}
\frac{\lambda_2(\Omega)}{\lambda_1(\Omega)} \le \frac{\lambda_2(\Omega^\star)}{\lambda_1(\Omega^\star)}, \quad \Omega
\subset \R^n, |\Omega^\star| = |\Omega|,
\end{equation}
where the ratio of the first two eigenvalues on $\Omega$ is compared to that on the spherical rearrangement
$\Omega^\star$. It has been pointed out in \cite{AB00} already, that an estimate in the form (\ref{EqMainResult}) is
the more natural result for settings where $\lambda_2/\lambda_1$ on a ball is a non-constant function of the ball's
radius. In the case of a domain on a hemisphere, for example, $\lambda_2/\lambda_1$ on a ball is an increasing function
of the radius \cite{AB00}. But by the Rayleigh-Faber-Krahn inequality for spheres \cite{S73} the radius of $S_1$ is
smaller than the one of the spherical rearrangement $\Omega^\star$. This means that an estimate in the form
(\ref{EqMainResult}), interpreted as
$$\frac{\lambda_2(\Omega)}{\lambda_1(\Omega)} \le \frac{\lambda_2(S_1)}{\lambda_1(S_1)}, \quad \Omega\subset \mathbb S^n,$$
is stronger than an inequality of the type (\ref{EqOriginalPPW}).

On the other hand, we have shown in \cite{BL06} that $\lambda_2/\lambda_1$ on balls is a strictly decreasing function
of the radius for the Dirichlet Laplacian in the space with a measure of inverted Gaussian density. In this case we can
apply an argument from \cite{AB00} to see that an estimate of the type (\ref{EqOriginalPPW}) cannot possibly hold true:
Consider a domain $\Omega$ that is constructed by attaching very long and thin tentacles to the ball $B$. Then the
first and second eigenvalues of the Laplacian on $\Omega$ are arbitrarily close to the ones on $B$. The spherical
rearrangement of $\Omega$ though can be considerably larger than $B$. This means that
$$\frac{\lambda_2(\Omega)}{\lambda_1(\Omega)} \approx \frac{\lambda_2(B)}{\lambda_1(B)} >
\frac{\lambda_2(\Omega^\star)}{\lambda_1(\Omega^\star)}, \qquad B,\Omega \subset (\R^n, e^{r^2} \di^n r)$$ %
clearly ruling out any inequality in the form of (\ref{EqOriginalPPW}).

With respect to the behavior of $\lambda_2/\lambda_1$ on balls, our present case of the hyperbolic space is similar to
the one of the inverted-Gaussian space:
\begin{Theorem}[Monotonicity of $\lambda_2/\lambda_1$ on balls in $\Hn$] \label{TheoremMonotonicity}
Let $\lambda_i(\theta,\rho)$ be the $i$-th eigenvalue of the Dirichlet-Laplacian on the ball of geodesic radius
$\theta$. Then $\lambda_2(\theta,\rho)/\lambda_1(\theta,\rho)$ is a strictly decreasing function of $\theta$.
\end{Theorem}
Repeating the argument given above for the inverted-Gaussian space, we see that an inequality of the type
(\ref{EqOriginalPPW}) can not hold true in $\Hn$.\\

Our proof of Theorem \ref{TheoremPPW} follows the general lines of \cite{AB92a}, though each step has to be modified to
fit our special setting. This is how the rest of the present article is organized:

In the following section we summarize briefly how $\Hn$ and the differential operators in it are defined in terms of
Riemannian geometry. For a comprehensive introduction to the spectral theory of differential operators on Riemannian
manifolds we refer the reader to \cite{C}.

In Section \ref{SectionIdentify} and Section \ref{SectionProofTheoremMonoton} we will state several results about the
first two Dirichlet eigenvalues on geodesic balls in $\Hn$. First, to identify the second eigenvalue, we prove a lemma
that is somewhat analogous to the Baumgartner-Grosse-Martin inequality \cite{BGM, BGM85, AB88} for Schr\"odinger
operators in Euclidean space. Second, we prove Theorem \ref{TheoremMonotonicity} and some other facts about the first
two eigenvalues on balls. In contrast to \cite{AB00} we do not apply perturbation theory to prove the monotonicity of
$\lambda_2/\lambda_1$ on balls of varying radius. Instead we use scaling properties of the eigenvalues in combination
with a lemma that is similar to Theorem \ref{TheoremPPW}, but that compares the eigenvalues on balls in hyperbolic
spaces of different curvature. We believe that this is the more natural way to prove Theorem \ref{TheoremMonotonicity}
and it has the advantage to deliver some additional lemmata as byproducts.

Section \ref{SectionProofTheoremPPW} shows the bare-bone structure of the proof of Theorem \ref{TheoremPPW}, which
consists as usual \cite{AB92a,AB00} in using the gap inequality for $\lambda_1$ and $\lambda_2$, the choice of suitable
test functions and finally the use of rearrangement techniques and monotonicity arguments.

The final four sections are to fill the gaps that are left open in Section \ref{SectionProofTheoremPPW}. First we prove
the ´center of mass' result for $\Hn$. Then, in sections \ref{SectionMonotonicityLemma} and \ref{SectionAnalysisOfZ} we
prove the monotonicity of certain functions $g$ and $B$. In comparison to the Euclidean or spherical case, this part of
our proof causes quite a lot of difficulties as can be seen in the corresponding sections.

Finally, in Section \ref{SectionChiti} a version of Chiti's comparison result is proven, in a slightly simplified way
compared to the proof in \cite{AB92a,AB00}.

\section{Preliminaries about the Laplacian operator in $\mathbb H^n$} \label{SectionPreliminaries}

In this section we briefly summarize how to define the hyperbolic space and the differential operators in it. We follow
the definitions of \cite{C}, where a more detailed discussion of the subject can be found. We fix a constant negative
curvature $\kappa < 0$ and set $\rho = (-\kappa)^{-1/2}$. Then we realize $\Hn$ as the ball $B^n(\rho) \subset \R^n$ of
radius $\rho$ endowed with the metric
\begin{equation} \label{EqMetric1}
ds^2 = \frac{4||dx||^2}{(1-||\vec x/\rho||^2)^2},
\end{equation}
where $||\cdot ||$ denotes the Euclidean norm. We define spherical coordinates by writing $\vec x \in B^n(\rho)$ as
$$\vec x = \rho \tanh(\theta/2\rho) \vec \chi,$$
where $\vec\chi \in \mathbb S^{n-1}$ is a vector of the unit sphere and $\theta \in [0,\infty)$. The hyperbolic metric
(\ref{EqMetric1}) takes in our new coordinates the form
\begin{equation} \label{EqMetric2}
ds^2 = d\theta^2 + \rho^2 \sinh^2 (\theta/\rho) \,\, ||d\vec\chi||^2.
\end{equation}
This metric induces an inner product $\langle.,.\rangle$ and norm $|\cdot|$ on the tangent space at any point of $\Hn$.
Each tangent space is isomorphic to $\R^n$ and its elements are associated with directional derivatives of $C^1$
functions defined on $\Hn$.

The Riemannian measure in $\Hn$ is given by
$$\di V = \rho^{n-1} \sinh^{n-1}(\theta/\rho)\, \di \theta \di \sigma,$$
where $\di\sigma$ is the measure on $\mathbb S^{n-1}$. We write $(.,.)$ for the scalar product in $L^2(\Hn)$.

We define the divergence of a vector field $X$ in a point $p$ by
$$\dvg X|_p = \textmd{trace}(\xi\rightarrow \nabla_\xi X),$$
where $\xi$ ranges over the tangent space in $p$ and $\nabla$ is the Levi-Civita connection of $\Hn$.

Given a function $f\in L^2(\Hn)$, we write $Y = \Grad f$ for a vector field $Y$ if
$$(Y,X) = -(f,\dvg X)$$
for all $C^1$ vector fields $X$ with compact support in $\Hn$. If $f$ is a $C^1$ function then
$$\langle \Grad f,\xi\rangle = \xi f$$
holds in each point $p$ of $\Hn$ and for each tangent vector $\xi$.

As usual, we define the Dirichlet Laplacian $-\Delta_\rho$ on some bounded domain $\Omega\subset\Hn$ by its quadratic
form
\begin{equation}\label{EqDefD}
D[u] = (\Grad u,\Grad u),
\end{equation}
defined on the completion of $C_0^\infty(\Omega)$ with respect to the norm induced by $D[u]$. It is known that
$-\Delta_\rho$ is then a positive self-adjoint operator with a pure point spectrum, and as mentioned before we call
$\lambda_i(\Omega,\rho)$ its $i$-th eigenvalue. In the case $\rho=1$ we will usually leave the second argument away.
Recall that in Theorem \ref{TheoremPPW} we defined $S_1$ to be a geodesic ball such that
$\lambda_1(\Omega,\rho)=\lambda_1(S_1,\rho)$. We chose $S_1$ to be centered at the origin of our spherical coordinate
system and we will call $\tmax$ the geodesic radius of $S_1$.

The differential expression of $-\Delta_\rho$ is
\begin{eqnarray}
-\Delta_\rho u &=& -\sinh^{1-n}(\theta/\rho) \, \frac{\partial}{\partial\theta}
\left(\sinh^{n-1}(\theta/\rho) \, \frac{\partial u}{\partial\theta}\right) \label{EqDifEx}\\
&&- \rho^{-2} \sinh^{-2}(\theta/\rho) \, \Delta_{\mathbb S^{n-1}} u,\nonumber
\end{eqnarray}
where $\Delta_{\mathbb S^{n-1}}$ is the Laplacian on the unit sphere.

The Rayleigh-Ritz characterization of the eigenvalues $\lambda_i$ of $-\Delta$ is completely analogous to the Euclidean
case: If $u\in L^2(\Omega)$ is some function in the domain of $D$ and if $u$ is orthogonal to the first $k-1$
eigenfunctions of $-\Delta$, then we have
$$\lambda_k(\Omega) \le \frac{D[u]}{(u,u)}.$$
This Rayleigh-Ritz formula is an important ingredient of our proof. Further, we need some more specific information on
the first two eigenvalues of $-\Delta$, which will be provided in the following section.

\section{Identifying the first two Dirichlet eigenvalues on geodesic balls in $\mathbb H^n$} \label{SectionIdentify}

A standard separation of variables (see, e.g., \cite{C}, p.40) shows that the eigenvalues and eigenfunctions of
$-\Delta_\rho$ on a geodesic ball in $\Hn$ of radius $\tmax$ are determined by the differential equation
\begin{equation}\label{EqDE}
-z''(\theta)-\frac{(n-1)}{\rho\tanh(\theta/\rho)} z'(\theta) +\frac{l\,(l+n-2)}{\rho^2 \sinh^2(\theta/\rho)} z(\theta)
= \lambda z(\theta),
\end{equation}
where $l=0,1,2,...$, with the boundary conditions $z'(0)=0$ (for $l=0$) or $z(\theta) \sim \theta^l$ as
$\theta\downarrow 0$ (for $l>0$) and $z(\tmax)=0$. Throughout the paper we will use the convention that $z_l(\theta)
> 0$ for $\theta \downarrow 0$. If one defines the left hand side of (\ref{EqDE}) as the operator $h_l$ applied to $z$ (with the boundary
conditions incorporated in the definition of the operator), then it is easy to see that $h_{l'} > h_l$ in the sense of
quadratic forms if $l'>l$. Consequently, the lowest eigenvalue $\lambda_1$ of $-\Delta_\rho$ on the geodesic ball is
$\lambda_1(h_0)$, while the second eigenvalue $\lambda_2$ on the geodesic ball must be either $\lambda_2(h_0)$ or
$\lambda_1(h_1)$. We show that the later is the case:
\begin{Lemma} \label{LemmaBGM}
The first eigenvalue of the Dirichlet Laplacian on a geodesic ball in $\Hn$ is the first eigenvalue of (\ref{EqDE})
with $l=0$, while the second eigenvalue on the geodesic ball is the first eigenvalue of (\ref{EqDE}) with $l=1$. The
second eigenvalue is $n$-fold degenerate.
\end{Lemma}
\begin{proof}
We will prove the lemma for $\rho = 1$. The general case follows directly from the behavior of the eigenvalues under
rescaling of $\rho$ (see Lemma \ref{LemmaScaling} below).

Assume that $z_l$ solves (\ref{EqDE}) for some fixed $\lambda$. Then one can verify that
$$z_l' - l\coth\theta \,z_l$$
satisfies (\ref{EqDE}) for $l$ replaced by $l+1$ and
$$z_l'+(l+n-2)\coth\theta\,z_l$$
satisfies (\ref{EqDE}) for $l$ replaced by $l-1$. From these facts and Frobenius theory it follows that
\begin{eqnarray}
z_{l+1} &=& -z_l' + l\coth\theta \, z_l, \label{EqZ1}\\
z_{l-1} &=& z_l' + (l+n-2)\coth\theta \, z_l. \label{EqZ2}
\end{eqnarray}
Putting $l=0$ in (\ref{EqZ1}) we get
\begin{equation}
z_1= -z_0'. \label{EqZ3}
\end{equation}
Setting $l=1$ in (\ref{EqZ2}) and multiplying with $\sinh^{n-1}\theta$ yields
\begin{equation}
\sinh^{n-1} \theta \, z_0 = (\sinh^{n-1}\theta \, z_1)'. \label{EqZ4}
\end{equation}
By Rolle's theorem, between any two zeros of $z_0$ there is a zero of $z_0'$, and hence of $z_1$ since (\ref{EqZ3})
holds. Similarly, between two zeros of $\sinh^{n-1}\theta \, z_1$ there is a zero of its derivative and hence of $z_0$
by (\ref{EqZ4}). Thus for fixed $\lambda>0$ the zeros of $z_0$ and $z_1$ on $[0,\infty)$ interlace.

Now consider $z_0$ and $z_1$ for $\lambda=\lambda_1(h_1)$. Then the first positive zero of $z_1$ is equal to the radius
$\tmax$ of our geodesic ball. It is clear by what we have just shown that $z_0$ has then exactly one zero in
$[0,\tmax]$ and the second zero of $z_0$ is greater than $\tmax$. This and the fact that the positive zeros of any
$z_l$ are decreasing functions of $\lambda$ show that $\lambda_2(h_0) > \lambda_1(h_1)$.

The degeneracy of the second eigenvalue on the geodesic ball follows from the details of the separation of variables.
\end{proof}

For future reference we state the following lemma.
\begin{Lemma} \label{LemmaMonotonicityZzero}
The solution $z_0$ of (\ref{EqDE}) with $l=0$ and some $\lambda >0$ is strictly decreasing on $[0,\theta_0]$, where
$\theta_0$ is the first positive zero of $z_0$. (Note that by our convention $z_0 > 0$.)
\end{Lemma}
\begin{proof}
The function $z_0$ satisfies
$$-(\sinh^{n-1}\theta\,z_0')' = \lambda \sinh^{n-1}\theta\,z_0 > 0 \quad \textmd{for }\theta\in[0,\theta_0],$$
which implies that $\sinh^{n-1}\theta\,z_0'$ is decreasing in $[0,\theta_0]$. Hence
$$\sinh^{n-1}\theta\,z_0' < (\sinh^{n-1}\theta\,z_0')|_{\theta=0} = 0,$$
which proves the Lemma.
\end{proof}

\section{Proof of Theorem \ref{TheoremMonotonicity}} \label{SectionProofTheoremMonoton}

To prove Theorem \ref{TheoremMonotonicity} we need the following lemmata which contain some basic information on the
behavior of the first two eigenvalues of $-\Delta_\rho$ on geodesic balls.

\begin{Lemma} \label{LemmaScaling}
Using the notation introduced in Section \ref{SectionPreliminaries}, we have for any $c > 0$, $\theta_0 > 0$ and $\rho
> 0$
\begin{equation}
\lambda_i(c\theta_0,c\rho) = c^{-2}\lambda_i(\theta_0,\rho).
\end{equation}
If $z_i(\theta)$ is the radial part of the $i$-th eigenfunction of $-\Delta_\rho$ on the ball of geodesic radius
$\theta_0$, then $z_i(\theta/c)$ is the radial part of the $i$--th eigenfunction of $-\Delta_{c\rho}$ on the ball of
radius $c\theta_0$.
\end{Lemma}
The proof of Lemma \ref{LemmaScaling} is straight-forward if one replaces $\theta$ in (\ref{EqDE}) by $\theta/c$ and
multiplies both sides with $c^2$ afterwards.

\begin{Lemma} \label{LemmaFirstEV}
The first eigenvalue $\lambda_1(\theta_0, \rho)$ of $-\Delta_\rho$ on a geodesic ball of radius $\theta_0$ is a
strictly decreasing function of $\theta_0$ and $\rho$, respectively.
\end{Lemma}

\begin{Lemma} \label{LemmaSecondEV}
Let $0<\rho_1<\rho_2$ and $\theta_1, \theta_2$ be such that
$$\lambda_1(\theta_1,\rho_1) = \lambda_1(\theta_2,\rho_2).$$
Then $\theta_1 > \theta_2$ and
$$\lambda_2(\theta_1,\rho_1) \le \lambda_2(\theta_2,\rho_2).$$
\end{Lemma}

\begin{proof}[Proof of Lemma \ref{LemmaFirstEV}]
By setting $z = (\rho\sinh(\theta/\rho))^{(1-n)/2} \, v$ one can check that for $l=0$ equation (\ref{EqDE}) is
associated with the one--dimensional Schr\"odinger operator
\begin{equation} \label{eq:3.15}
H_{0}=-\frac{d^2}{d\theta^2}+ \frac{n-1}{4} \left(\frac{n-1}{\rho^2}+\frac{n-3}{\rho^2\sinh^2(\theta/\rho)}\right),
\end{equation}
acting on $L^2((0,\theta_0),d\theta)$ with Dirichlet boundary conditions imposed at $0$ and $\theta_0$. We rewrite the
potential term of $H_0$ as
$$V =  \frac{(n-1)^2}{4 \rho^2 \tanh^2(\theta/\rho)} - \frac{2}{\rho^2\sinh^2(\theta/\rho)},$$
where it can be checked easily that each of the two summands is strictly decreasing in $\rho$. Thus
$\lambda_1(\theta_0,\rho)$ is a strictly decreasing function of $\rho$ and, by domain monotonicity, in $\theta_0$.
\end{proof}

\begin{proof}[Proof of Lemma \ref{LemmaSecondEV}]
Let $B_{1}$ and $B_2$ be balls of geodesic radii $\theta_1, \theta_2$ in the hyperbolic spaces of curvature $\kappa_{1}
= -\rho_1^{-2}$ and $\kappa_2 = -\rho_2^{-2}$, respectively. We call $y_0(\theta)$ the first eigenfunction of
$-\Delta_{\rho_1}$ on $B_1$, $z_0(\theta)$ the first eigenfunction of $-\Delta_{\rho_2}$ on $B_2$ and $z_1(\theta)$ the
radial part of the second eigenfunction of $-\Delta_{\rho_2}$ on $B_2$. Keep in mind Lemma \ref{LemmaBGM} and our
convention to choose eigenfunctions like $y_0, z_0$ and $z_1$ positive. We will assume $y_0$ and $z_0$ to be normalized
to one in $L^2(B_1, \di V_1)$ and $L_2(B_2, \di V_2)$, respectively. By $\di V_1$ and $\di V_2$ we denote the
hyperbolic measures corresponding to the curvatures $\kappa_1$ and $\kappa_2$.

Because $\rho_1<\rho_2$ and because the first eigenvalues on $B_1$ and $B_2$ are equal, we can conclude from Lemma
\ref{LemmaFirstEV} that $\theta_2 < \theta_1$.

In the same way as below in Section \ref{SectionProofTheoremPPW}, with the only difference that $\rho$ is arbitrary and
that the `center of mass' theorem is trivial by symmetry, we derive the gap formula
\begin{equation}\label{EqGapBall}
\lambda_2(\theta_1,\rho_1)-\lambda_1(\theta_1,\rho_1) \le \frac{\int_{B_1} y_0^2(\theta) B_1(\theta) \di
V_1}{\int_{B_1} y_0^2(\theta) g(\theta)^2 \di V_1}
\end{equation}
with
\begin{equation} \label{EqDefBrho}
B_1(\theta) = \left(\frac{\partial g}{\partial \theta}\right)^2 +
\frac{n-1}{\rho_1^2\sinh^2(\theta/\rho_1)}g^2(\theta).
\end{equation}
We chose $g(\theta)$ to be
\begin{equation}
g(\theta) = \left\{\begin{array}{ll} \frac{z_1(\theta)}{z_0(\theta)} & \textmd{for } \theta \in [0,\theta_2),\cr
\lim_{\theta\uparrow\theta_2} g(\theta) & \textmd{for } \theta \in [\theta_2,\theta_1). \end{array}\right.
\end{equation}
It will be proven in Section \ref{SectionMonotonicityLemma} that $g$ is an increasing function if $\rho_1 = 1$. Using
Lemma \ref{LemmaScaling} one can see easily that $g$ is then increasing for any $\rho_1>0$. In almost the same way as
in Section \ref{SectionMonotonicityLemma} one can then also show that $B_1$ is decreasing for any $\rho_1>0$.

In what follows it turns out to be more convenient to write (\ref{EqGapBall}) as
\begin{equation}\label{EqGapBall2}
\lambda_2(\theta_1,\rho_1)-\lambda_1(\theta_1,\rho_1) \le \frac{\int_0^{\theta_1} y_0^2(\theta) B_1(\theta)
A_{\rho_1}'(\theta) \di \theta}{\int_0^{\theta_1}  y_0^2(\theta) g(\theta)^2 A_{\rho_1}'(\theta) \di \theta}
\end{equation}
where
$$A_\rho(\theta) = nC_n \int_0^\theta \rho^{n-1} \sinh^{n-1} (\theta'/\rho) \di \theta'$$
is the measure of a ball of geodesic radius $\theta$ and curvature $\kappa = -\rho^2$.

\begin{Fact} \label{FactIntersections}
For any $\gamma > 0$ the functions $y_0(\theta)$ and $\gamma z_0(\theta)$ do not intersect more than once on
$[0,\theta_1]$ (setting $z_0(\theta) = 0$ for $\theta>\theta_2$).
\end{Fact}
\begin{proof}
We consider the functions
$$p_1(\theta) = \frac{y'_0 (\theta)}{y(\theta)} \qquad \text{and}\qquad p_2(\theta) = \frac{z'_0(\theta)}{z_0(\theta)}.$$
Let us assume that $y_0$ and $\gamma z_0$ intersect more than once. Then, because both functions are positive and
continuous, at one of these intersections (let's say at $\theta=\theta_3$) we have  $p_2(\theta_3) \ge p_1(\theta_3)$.
At $\theta_2$ the function $p_2$ goes to minus infinity, while $p_1$ remains finite. This means that there must be some
$\theta_4 \in [\theta_3,\theta_2)$ such that $p_2(\theta_4)=p_1(\theta_4)$ and $p'_2(\theta_4) \le p'_1(\theta_4)$

Using (\ref{EqDE}) we can derive the Riccati equation
\begin{equation*}
p_i'(\theta) = -p_i^2(\theta) - \frac{n-1}{\rho_i \tanh(\theta/\rho_i)} p_i(\theta) - \lambda_1 \qquad (i = 1,2).
\end{equation*}
Evaluated at $\theta = \theta_4$, this equation tells us that
\begin{eqnarray*}
p'_2(\theta_4) - p'_1(\theta_4) &=& p_1(\theta_4) (n-1) \left(\frac{1}{\rho_1 \tanh (\theta/\rho_1)} - \frac{1}{\rho_2
\coth(\theta/\rho_2)}\right) \\
&>& 0
\end{eqnarray*}
which is a contradiction to what has been said before and thus proves the lemma.
\end{proof}

\begin{Fact} \label{FactIntersections2}
There is a $\theta_5 \in (0,\theta_2)$ such that
\begin{equation} \label{EqF1}
\begin{array}{ll}
z_0(\theta) A_{\rho_2}'(\theta) \ge y_0(\theta) A_{\rho_1}'(\theta) \quad\quad& \textmd{for } 0 < \theta < \theta_5,\cr%
z_0(\theta) A_{\rho_2}'(\theta) \le y_0(\theta) A_{\rho_1}'(\theta) \quad\quad& \textmd{for } \theta_5 < \theta < \theta_1.\cr%
\end{array}
\end{equation}
\end{Fact}
\begin{proof}
By the normalization of $y_0$ and $z_0$ it is clear that the two functions $z_0 A_{\rho_2}'$ and $y_0 A_{\rho_1}'$
intersect at least once on $(0,\theta_2)$. This means that also $z_0 A_{\rho_2}'/A_{\rho_1}'$ and $y_0$ intersect at
least once, say in $\theta_5$. Then by Fact \ref{FactIntersections} the functions
$$z_0(\theta) A_{\rho_2}'(\theta_5)/ A_{\rho_1}'(\theta_5) \qquad \textmd{and}\qquad  y_0(\theta)$$
intersect exactly once on $(0,\theta_1)$, and because $y_0(\theta_2) > z_0(\theta_2) = 0$ we know that
\begin{equation} \label{EqF2}
\begin{array}{ll}
z_0(\theta) A_{\rho_2}'(\theta_5)/ A_{\rho_1}'(\theta_5)\ge y_0(\theta) \quad\quad& \textmd{for } 0<\theta<\theta_5,\cr%
z_0(\theta) A_{\rho_2}'(\theta_5)/ A_{\rho_1}'(\theta_5)\le y_0(\theta) \quad\quad& \textmd{for } \theta_5<\theta<\theta_1.\cr%
\end{array}
\end{equation}
Finally we note that
$$A_{\rho_2}'/ A_{\rho_1}' = \left(\frac{\rho_2 \sinh(\theta/\rho_2)}{\rho_1\sinh(\theta/\rho_1)}\right)^{n-1}$$
is a decreasing function, such that (\ref{EqF1}) follows from (\ref{EqF2}).
\end{proof}

Using Fact \ref{FactIntersections2} and the monotonicity properties of $g$ and $B$, we can derive from
(\ref{EqGapBall2}) that
\begin{equation}\label{EqGapBall3}
\lambda_2(\theta_1,\rho_1)-\lambda_1(\theta_1,\rho_1) \le \frac{\int_0^{\theta_1} z_0^2(\theta) B_1(\theta)
A_{\rho_2}'(\theta) \di \theta}{\int_0^{\theta_1}  z_0^2(\theta) g(\theta)^2 A_{\rho_2}'(\theta) \di \theta}
\end{equation}
If we define $B_2$ in the same way as $B_1$ in (\ref{EqDefBrho}), just replacing $\rho_1$ by $\rho_2$, we have
$B_2(\theta) > B_1(\theta)$ and therefore
\begin{equation}\label{EqGapBall4}
\lambda_2(\theta_1,\rho_1)-\lambda_1(\theta_1,\rho_1) \le \frac{\int_0^{\theta_1} z_0^2(\theta) B_2(\theta)
A_{\rho_2}'(\theta) \di \theta}{\int_0^{\theta_1}  z_0^2(\theta) g(\theta)^2 A_{\rho_2}'(\theta) \di \theta} =
\lambda_2(\theta_2,\rho_2)-\lambda_1(\theta_2,\rho_2),
\end{equation}
proving Lemma \ref{LemmaSecondEV}.
\end{proof}

\begin{proof}[Proof of Theorem \ref{TheoremMonotonicity}]
We fix some $0< \theta_1 < \theta_2$ and some $\rho > 0$ and define the function $\Theta(c)$ for $0 < c \le 1$
implicitly by
\begin{equation} \label{EqTheta}
\lambda_1(\Theta(c), \rho) = \lambda_1(c \theta_2, c\rho).
\end{equation}
Then $\Theta$ is well defined (by Lemma \ref{LemmaFirstEV}) and continuous, $\Theta(1) = \theta_2$ and
$\Theta(c\rightarrow 0) = 0$. We can therefore chose $c_0 \in (0,1)$ such that $\Theta(c_0) = \theta_1$. Theorem
\ref{TheoremMonotonicity} is now proven by the following chain of (in)equalities:
\begin{eqnarray*}
\frac{\lambda_2(\theta_1,\rho)}{\lambda_1(\theta_1,\rho)} &=&
\frac{\lambda_2(\Theta(c_0),\rho)}{\lambda_1(\Theta(c_0),\rho)} \\
&\ge& \frac{\lambda_2(c_0\theta_2,c_0\rho)}{\lambda_1(c_0\theta_2,c_0\rho)}
=\frac{c_0^2\lambda_2(\theta_2,\rho)}{c_0^2\lambda_1(\theta_2,\rho)} =
\frac{\lambda_2(\theta_2,\rho)}{\lambda_1(\theta_2,\rho)}
\end{eqnarray*}
Here the first step is just the definition of $c_0$, the second step follows from Lemma \ref{LemmaSecondEV} and the
third one from Lemma \ref{LemmaScaling}.
\end{proof}

\section{Proof of Theorem \ref{TheoremPPW}} \label{SectionProofTheoremPPW}

\begin{proof}[Proof of Theorem \ref{TheoremPPW}]
First we argue that it is sufficient to prove the theorem for the case $\rho = 1$: Using the differential expression
(\ref{EqDifEx}) of $-\Delta_\rho$ one can check that if $u(\theta,\vec\chi)$ is some eigenfunction of $-\Delta_\rho$
corresponding to the eigenvalue $\lambda$, then $\tilde u(\theta,\vec\chi) = u(\rho\,\theta,\vec\chi)$ is an
eigenfunction of $-\Delta_1$ corresponding to the eigenvalue $\tilde\lambda = \lambda\rho^2$. One can convince oneself
that it is therefore sufficient to prove Theorem \ref{TheoremPPW} for $\rho = 1$.

We call $u_1$ the first eigenfunction of $-\Delta$ on $\Omega$. Further we recall that $S_1$ is a geodesic ball in
$\Hn$ such that $\lambda_1(\Omega) = \lambda_1(S_1)$ and we write $z_0(\theta)$ for the corresponding eigenfunction. By
Lemma \ref{LemmaBGM} the second eigenvalue $\lambda_2(S_1)$ is $n$-fold degenerate and the corresponding eigenspace is
spanned by the functions $z_1(\theta) \chi_k$, where $z_1$ is the solution of (\ref{EqDE}) for $l=1$.

Let $P\neq 0$ be a function on $\Omega$ such that that $Pu_1$ is in the domain of $D$ as defined in (\ref{EqDefD}) and
\begin{equation} \label{EqConditionP}
\int_\Omega Pu_1^2 \di V =0.
\end{equation}
Then by the Rayleigh-Ritz theorem we have the estimate
\begin{eqnarray*}
\lambda_2(\Omega)-\lambda_1(\Omega) &=& \frac{D[Pu_1]}{(Pu_1,Pu_1)} - \lambda_1\\
&=& (Pu_1,Pu_1)^{-1} \int_\Omega \left( \langle \Grad Pu_1,\Grad Pu_1 \rangle - \lambda_1 P^2u_1^2\right) \di V
\end{eqnarray*}
Using (\ref{EqDefD}) and the $\Hn$ generalizations of well known formulas for differential operators in the Euclidean
space \cite{C}, we find the gap inequality
\begin{equation}\label{EqGap}
\lambda_2(\Omega) - \lambda_1(\Omega) \le\frac{\int_\Omega u_1^2 \langle \Grad P,\Grad P\rangle \di V}{\int_\Omega P^2
u_1^2 \di V}.
\end{equation}
We do not only chose only one function $P$ but rather a set of $n$ functions, namely
$$P_i(\theta, \vec \chi) := g(\theta) \chi_i$$
with
\begin{equation} \label{EqDefg}
g(\theta) = \left\{\begin{array}{ll} \frac{z_1(\theta)}{z_0(\theta)} & \textmd{for } \theta \in [0,\tmax),\cr
\lim_{\theta\uparrow\tmax} g(\theta) & \textmd{for } \theta \ge \tmax. \end{array}\right.
\end{equation}
Recall that $\tmax$ is the geodesic radius of $S_1$ and that by convention $z_0$ and $z_1$ are positive. With this
choice of the $P_i$ and by our `center of mass' theorem (see Section \ref{SectionCOM}) it is always possible to shift
$\Omega$ in $\Hn$ such that the condition (\ref{EqConditionP}) is met.

We calculate that
\begin{eqnarray*}
\sum_{i=1}^n P_i^2(\theta,\vec\chi) &=& g^2(\theta),\\
\sum_{i=1}^n \langle\Grad P_i,\Grad P_i \rangle &=& \left(\frac{\partial g}{\partial \theta}\right)^2 + (n-1)
\sinh^{-2}(\theta)\, g^2(\theta).
\end{eqnarray*}
We multiply inequality (\ref{EqGap}) by $\int_\Omega P^2 u_1^2 \di V$ and then sum up over $i = 1,...,n$ to obtain
\begin{equation} \label{EqGap2}
\lambda_2(\Omega)-\lambda_1(\Omega) \le \frac{\int_\Omega u_1^2(\theta,\vec\chi) B(\theta) \di V}{\int_\Omega
u_1^2(\theta,\vec\chi) g^2(\theta) \di V}
\end{equation}
with
\begin{equation} \label{EqDefB}
B(\theta) = g'(\theta)^2+ (n-1) \sinh^{-2}(\theta)\, g^2(\theta).
\end{equation}
To conclude the proof of Theorem \ref{TheoremPPW} we need the following two chains of inequalities:
\begin{eqnarray} \label{EqChain1}
\int_\Omega u_1^2(\theta,\vec\chi)  B(\theta) \di V &\le& \int_{\Omega^\star} {u_1^\star}(\theta)^2 B^\star(\theta) \di V \\
&\le& \int_{\Omega^\star} {u_1^\star}(\theta)^2 B(\theta) \di V \le \int_{S_1} z_0^2(\theta) B(\theta) \di V \nonumber
\end{eqnarray}
and
\begin{eqnarray} \label{EqChain2}
\int_\Omega u_1^2(\theta,\vec\chi) g(\theta)^2 \di V &\ge& \int_{\Omega^\star} {u_1^\star}(\theta)^2 g_\star(\theta)^2 \di V\\
&\ge& \int_{\Omega^\star} {u_1^\star}(\theta)^2 g(\theta)^2 \di V \ge \int_{S_1} z_0^2(\theta) g^2(\theta) \di
V\nonumber
\end{eqnarray}
Here we assume that $z_0$ is normalized such that $\int_\Omega u_1^2 \di V = \int_{S_1} z_0^2 \di V$. With a star we
denote the spherical rearrangements of functions, i.e., $f^\star$ is the spherical decreasing rearrangement of some
function $f$ and $f_\star$ the increasing one (for more details see Section \ref{SectionChiti}). In each of
(\ref{EqChain1}) and (\ref{EqChain2}) the first inequality follows directly from the properties of rearrangements
\cite{T}. The second inequality is true because of the monotonicity properties of $g$ and $B$, that will be proven in
Section \ref{SectionMonotonicityLemma} below. The third inequality finally follows from our version of Chiti's
comparison result that will be presented in Section \ref{SectionChiti}, and from the monotonicity properties of $g$ and
$B$ again. Finally, from (\ref{EqGap2}), (\ref{EqChain1}) and (\ref{EqChain2}) we obtain
\begin{equation}
\lambda_2(\Omega) - \lambda_1(\Omega) \le \frac{\int_{S_1} z_0^2(\theta) B(\theta) \di V}{\int_{S_1}z_0^2(\theta)
g^2(\theta) \di V} = \lambda_2(\Omega) - \lambda_1(S_1).
\end{equation}
With $\lambda_1(\Omega) = \lambda_1(S_1)$ this yields
\begin{equation}\label{EqFinal}
\lambda_2(\Omega) \le \lambda_2(S_1),
\end{equation}
proving Theorem \ref{TheoremPPW}. (It is easy to check that equality in (\ref{EqFinal}) holds if and only if $\Omega$
itself is a geodesic ball.)
\end{proof}

\section{`Center of mass' result for domains in $\mathbb H^n$} \label{SectionCOM}
In this section we will show that $\Omega$ can always be shifted in $\Hn$ such that the condition (\ref{EqConditionP})
is fulfilled. Let us first make a remark though on the notion of `shifting' in $\Hn$. It is well known (see, e.g.,
\cite{C}, p. 37) that the metric of $\Hn$ has always the form (\ref{EqMetric2}), no matter which point of $\Hn$ we
chose as the origin of our coordinate system. This can most easily be seen in the Minkovski space realization of $\Hn$:
We define
$$\tHn := \{\vec y \in \R^n_1:  y_1^2+\dots +y_n^2-y_{n+1}^2= -1\}$$
and endow it with the restriction of the Minkovski metric in $\R^n_1$ to $\tHn$. This makes $\Hn$ and $\tHn$ isometric
spaces and an example of an isometry between them is
$$I: \Hn \rightarrow \tHn, \, \vec x = \tanh \frac \theta 2 \vec \chi \rightarrow \vec y = (\sinh \theta \vec
\chi, \cosh \theta).$$%
It is further known that each Lorentz transformation in $\R^n_1$ induces an isometry of $\tHn$ onto itself, the
hyperbolic analog of shifts and rotations in the Euclidean space. The group of these transformations has the
transitivity property, i.e., for any two points $p_1,p_2\in\tHn$ there exists a Lorentz transformation that maps $p_1$
on $p_2$.

Every Lorentz transformation $R$ on $\tHn$ induces an isometry $IRI^{-1}$ on $\Hn$. Especially, if we subject our
domain $\Omega$ to such a transformation (which we call a `shift' of $\Omega$), the eigenvalues of the Laplcian don't
change.

\begin{Theorem}[`Center of Mass'] \label{TheoremCOM}
Let $g(\theta)$ be a positive continuous function on $[0,\infty)$ and $P_i(\vec x) = \chi_i g(\theta)$. Then one can
shift $\Omega$ (and $u_1$ with it) such that
$$\int_\Omega P_i(\theta,\vec\chi) u_1^2(\theta,\vec \chi) \di V = 0 \quad \textmd{for all }i=1,\dots,n.$$
\end{Theorem}

\begin{proof}
We will prove Theorem \ref{TheoremCOM} in the Minkovski space representation of $\Hn$. Let $\tilde\Omega = I(\Omega)$,
$\tilde u_1(\vec y) = u_1(I^{-1}(\vec y))$ and $\tilde P_i(\vec y) = P_i(I^{-1}(\vec y))$. Then
$$\int_\Omega P_i(\vec x) u_1^2(\vec x) \di V = \int_{\tilde\Omega} \tilde P_i(\vec y) \tilde u_1^2(\vec y) \di
\tilde V.$$%
Because it doesn't matter whether we shift $\Omega$ and $u_1$ or $P_i$, we only have to show that there is some Lorentz
transformation $R$ such that
$$\int_{\tilde\Omega} \tilde P_i(R \vec y) \tilde u_1^2(\vec y) \di \tilde V = 0.$$
Let $\Theta(\vec z,\vec y)$ be the $\theta$ coordinate of $\vec y$ after a Lorentz transformation that maps $\vec z$ to
$(0,\dots,0,1)$ and set
$$\vec v(\vec z) = \int_{\tilde\Omega} \frac{\vec y}{\sinh\Theta(\vec z,\vec y)} g(\Theta(\vec z,\vec y)) \tilde u_1^2
\di \tilde V,$$ %
where the integration variable is $\vec y$. We write $\Pi: \tHn \rightarrow \R^n$ for the projection
$$\Pi \vec y = (y_1,\dots, y_n) \quad \textmd{for }\vec y \in \tHn.$$
Assume that there is a $\vec z_0 \in \tHn$ and $\alpha \in \R$ such that
\begin{equation}\label{EqCOM2}
\vec v(\vec z_0) = \alpha \vec z_0.
\end{equation}
Under this assumption we chose $R$ to be a Lorentz transformation that maps $\vec z_0$ to $(0,\dots,0,1)$. Then the
$\theta$-coordinate of $R\vec y$ is $\Theta(\vec z_0,\vec y)$ and the $\vec\chi$-coordinate is $\Pi R\vec y / \sinh
\Theta(\vec z_0,\vec y)$, such that
\begin{eqnarray*}
\int_{\tilde\Omega} \tilde P_i(R \vec y) \tilde u_1^2(\vec y) \di \tilde V &=& \int_{\tilde\Omega} \frac{(\Pi R \vec
y)_i}{\sinh \Theta(\vec z_0,\vec y)} g(\Theta(\vec z_0,\vec y))\, \tilde u_1^2(\vec y) \di \tilde V\\
&=& \bigl(\Pi R\vec v(\vec z_0)\bigr)_i\\
&=& \alpha (\Pi R \vec z_0)_i = 0
\end{eqnarray*}
Thus it only remains to show that a $\vec z_0$ exists that meets condition (\ref{EqCOM2}). The projection $\Pi$ has a
well defined inverse
$$\Pi^{-1} \vec \xi = (\xi_1,\dots,\xi_n, \sqrt{1+\xi_1^2+\dots+\xi_n^2}) \quad \textmd{for }\vec\xi \in \R^n.$$
Also $\Pi\tilde\Omega \subset \R^n$ is a bounded domain and we can chose a ball $B_R \subset \R^n$, centered at the
origin and of Euclidean radius $R$, such that $\Pi\tilde\Omega$ is contained in $B_R$. On $B_R$ we define the vector
field $\vec w: B_R \rightarrow \R^n$ with
\begin{equation}\label{EqCOM3}
\vec w(\vec \xi) = \Pi \left(\vec v(\Pi^{-1}\vec \xi) - \frac{[\vec v(\Pi^{-1} \vec\xi)]_{n+1}}{[\Pi^{-1}\vec
\xi]_{n+1}} \Pi^{-1}\vec\xi\right).
\end{equation}
The set $\Pi^{-1} B_R$ and the origin of $\R^n_1$ span the conus
$$C = \left\{\vec y \in \R^n_1 : y_{n+1} \ge 0 \textmd{ and }\frac{y_1^2+\dots+y_n^2}{y_{n+1}^2} \le \frac{R^2}{R^2+1}\right\}$$
and $\tilde\Omega$ is contained in $C$. Therefore also $\vec v(\vec z)$, being an integral over vectors in
$\tilde\Omega$ with positive coefficients, lies in this conus for every $\vec z \in \tHn$. Thus
\begin{equation}\label{EqCOM1}
\frac{||\Pi\vec v(\vec z)||}{[\vec v(\vec z)]_{n+1}} \le \frac{R}{\sqrt{R^2+1}}.
\end{equation}
Consequently, for any $\vec \xi \in \partial B_R$ we have, using first the Cauchy-Schwarz inequality and then
(\ref{EqCOM1}),
\begin{eqnarray*}
\vec w(\vec \xi) \cdot \vec \xi &=& \vec v(\Pi^{-1}\vec\xi)\cdot\vec\xi - \frac{[\vec v(\Pi^{-1}
\vec\xi)]_{n+1}}{[\Pi^{-1}\vec \xi]_{n+1}} ||\vec\xi||^2\\
&\le& ||\Pi\vec v(\Pi^{-1}\vec \xi)|| \, R - \frac{[\vec v(\Pi^{-1}
\vec\xi)]_{n+1}}{\sqrt{R^2+1}} R^2\\
&\le&  \frac{[\vec v(\Pi^{-1}\vec \xi)]_{n+1}}{\sqrt{R^2+1}} R^2 - \frac{[\vec v(\Pi^{-1}\vec
\xi)]_{n+1}}{\sqrt{R^2+1}} R^2 = 0
\end{eqnarray*}
This means that $\vec w(\vec \xi)$ points inward everywhere on $\partial B_R$. Then by the Brouwer Fixed Point Theorem
(see \cite{M}, p. 369, prob. 7 d)) there must be some point $\vec \xi_0 \in B_R$ such that $\vec w(\vec \xi_0) = 0$. We
put $\vec\xi_0$ for $\vec\xi$ into (\ref{EqCOM3}) and multiply with $\Pi^{-1}$ to get
$$\vec v(\Pi^{-1}\vec \xi_0) = \frac{[\vec v(\Pi^{-1} \vec\xi_0)]_{n+1}}{[\Pi^{-1}\vec \xi_0]_{n+1}} \Pi^{-1}\vec\xi_0.$$
Thus, setting $\vec z_0 = \Pi^{-1}\vec \xi_0$ we have proven the existence of a $\vec z_0$ that meets condition
(\ref{EqCOM2}).
\end{proof}

\section{A monotonicity lemma} \label{SectionMonotonicityLemma}
This section is devoted to the proof of the following lemma that states the monotonicity properties of $B$ and $g$ and
is crucial for the proof of the PPW conjecture in $\mathbb H^n$.
\begin{Lemma}[Monotonicity of $B$ and $g$]\label{LemmaMonotonicity}
The function $g(\theta)$ defined in (\ref{EqDefg}) is increasing and the function $B(\theta)$ defined in (\ref{EqDefB})
is decreasing.
\end{Lemma}
Similar to the proceeding in \cite{AB00} we will prove Lemma \ref{LemmaMonotonicity} by analyzing the function
\begin{equation} \label{EqDefq}
q(\theta) = \frac{\theta g'(\theta)}{g(\theta)} = \theta\left(\frac{z_1'}{z_1}-\frac{z_0'}{z_0}\right),
\end{equation}
which is defined by (\ref{EqDefq}) for $\theta \in (0,\tmax)$ only, but can be extended to a continuous and
differentiable function on the whole interval $[0,\tmax]$. We will show that $q(\theta)$ has the properties
\begin{eqnarray}
q(\theta) &\ge& 0, \label{Eqq1}\\
q(\theta) &\le& 1 \quad \textmd{and}\label{Eqq2}\\
q'(\theta) &\le& 0\label{Eqq3}
\end{eqnarray}
for all $\theta \in [0,\tmax]$. From (\ref{Eqq1}), the definition (\ref{EqDefq}) of $q$ and the fact that $g \ge 0$ we
conclude directly that $g' \ge 0$ and thus $g$ is increasing. Moreover, by solving (\ref{EqDefq}) for $g'$,
differentiating both sides of the resulting equation and putting (\ref{EqDefq}) in again, we get
$$g''(\theta) = \frac{g}{\theta^2}(\theta q'+q(q-1)).$$
Thus we can conclude from (\ref{Eqq1}), (\ref{Eqq2}) and (\ref{Eqq3}) that $g'' \le 0$. For the function
$$B(\theta) = g'(\theta)^2 + \frac{\nu g^2(\theta)}{\sinh^2(\theta)}$$
this means that the first summand $g'(\theta)^2$ is decreasing. But then also the second summand is decreasing: Because
$g(0)=0$ and because $g'$ is decreasing, we have the estimate
\begin{equation}
g(\theta)= \int_0^\theta g'(\tau) \di \tau \ge \int_0^\theta g'(\theta) \di \tau = g'(\theta)\, \theta.
\end{equation}
Thus we can calculate
\begin{eqnarray*}
\left(\frac{g(\theta)}{\sinh\theta}\right)' &=&
\frac{1}{\sinh\theta}\left(g'(\theta)-\frac{g(\theta)}{\tanh\theta}\right)\\
&\le& \frac{1}{\sinh\theta}\left(g'(\theta)-\frac{g'(\theta)\,\theta}{\tanh\theta}\right)\\
&=&\frac{g'(\theta)}{\sinh\theta} (1-\theta/\tanh\theta)
\end{eqnarray*}
This is negative because $g'$ is positive and $\theta\ge\tanh\theta$. This means that the second summand in the
definition of $B$ also decreases.

\begin{Remark}
In \cite{AB00} the function $q$ has been defined as in (\ref{EqDefq}), but with $\sin \theta$ replacing the factor
$\theta$. One might wonder whether in our $\mathbb H^n$ case the function $q$ should be defined rather with a factor
$\sinh\theta$ than just $\theta$. Actually this would simplify the Riccati equation for $q$ that we will derive below.
It turns out, however, that in this case $q$ would not be smaller than one anymore, destroying our scheme of the proof.
We thus have to stick to our definition (\ref{EqDefq}) of $q$ and deal with lengthened calculus below.
\end{Remark}

We have seen that Lemma \ref{LemmaMonotonicity} can be shown by proving the equations (\ref{Eqq1}), (\ref{Eqq2}) and
(\ref{Eqq3}), so this is what we will do in the remainder of this section. We begin with some basic properties of $q$.
From (\ref{EqDefq}) we can see quickly that
\begin{equation} \label{EqBoundaryq}
q(0) = 1,\quad q'(0) = 0 \quad \textmd{and} \quad q(\tmax) = 0.
\end{equation}
Differentiating (\ref{EqDefq}) and replacing the second derivatives of $z_0$ and $z_1$ according to the differential
equations they fulfill (equation (\ref{EqDE}) for $l=0, \lambda=\lambda_1(S_1)$ and $l=1, \lambda=\lambda_2(S_1)$,
respectively), we derive the Riccati equation
\begin{equation}\label{EqRiccatiq}
q' = \frac{q(1-q)}{\theta} - \nu q \coth \theta + \frac{\nu\theta}{\sinh^2\theta} - \theta \epsilon - 2pq.
\end{equation}
Here we have defined
$$p(\theta) = z_0'(\theta) / z_0(\theta),\quad \epsilon = \lambda_2-\lambda_1 \quad \textmd{and} \quad \nu = n-1.$$
We can also establish a Riccati equation for $p$ in a similar fashion as we derived (\ref{EqRiccatiq}). The result is
\begin{equation}\label{EqRiccatip}
p' = -p^2 - \nu p \coth \theta - \lambda_1.
\end{equation}
Now we set
\begin{equation} \label{EqDefT}
T(\theta,y) :=  \frac{y(1-y)}{\theta} - \nu y \coth \theta + \frac{\nu\theta}{\sinh^2\theta} - \theta \epsilon -
2p(\theta)y,
\end{equation}
for $y \in \R$ and $\theta \in (0,\tmax)$. Then by (\ref{EqRiccatiq}) we have
\begin{equation}\label{EqqT}
q'(\theta) = T(\theta,q(\theta)) \quad \textmd{for all}\quad \theta\in (0,\tmax).
\end{equation}
We call $T'$ the derivative of $T$ by $\theta$ and calculate
\begin{equation} \label{EqTprime}
T'(\theta,y) = \frac{y^2-y}{\theta^2} + \frac{(y+1)\nu}{\sinh^2 \theta} -\epsilon - \frac{2\nu\theta\cosh
\theta}{\sinh^3 \theta} - 2p'(\theta) y.
\end{equation}
We are interested in the behavior of $T'$ at points $(\theta, y)$ where $T=0$. Thus we first use (\ref{EqDefT}) to
calculate the value of $p$ at such points:
\begin{equation} \label{EqpatTzero}
p_{T=0} = \frac{1}{2y}\left( \frac{y(1-y)}{\theta} - \nu y \coth \theta - \theta \epsilon  +
\frac{\nu\theta}{\sinh^2\theta}\right).
\end{equation}
Then we put the Riccati equation (\ref{EqRiccatip}) for $p'$ into (\ref{EqTprime}) and finally eliminate $p$ with the
help of (\ref{EqpatTzero}). The result is
\begin{eqnarray}\label{EqTprimeatTzero}
T'_{T=0} &=& \frac{y^2-y}{\theta^2} + \frac{(y+1)\nu}{\sinh^2 \theta} -\epsilon - \frac{2\nu\theta\cosh \theta}{\sinh^3
\theta} +2y\lambda_1\nonumber\\
&&+\frac{1}{2y}\left(\frac{y-y^2}{\theta}-y\nu\coth\theta-\epsilon\theta+\frac{\nu\theta}{\sinh^2\theta}\right)^2\\
&& + \nu \coth \theta \left(\frac{y-y^2}{\theta}-y\nu\coth\theta
-\epsilon\theta+\frac{\nu\theta}{\sinh^2\theta}\right).\nonumber
\end{eqnarray}
The right hand side of (\ref{EqTprimeatTzero}) we be called $Z_y(\theta)$. The above analysis holds for all $\theta \in
(0,\tmax)$. For $y=1$ the limits of $Z_y$ and $T'$ as $\theta\rightarrow 0$ are finite and one can calculate from
(\ref{EqRiccatip}), (\ref{EqTprime}) and (\ref{EqTprimeatTzero}) that
\begin{eqnarray*}
p'(0) &=&\lim_{\theta\rightarrow 0} p'(\theta) = - \frac{\lambda_1}{\nu+1} \\
T'(0,1) &=& \lim_{\theta\rightarrow 0} T'(\theta,1) = -\lambda_2 + \left(1+\frac{2}{\nu+1}\right)\lambda_1
-\frac 23
\nu,\\
Z_1(0) &=& \lim_{\theta\rightarrow 0} Z_1(\theta) = (\nu+1) \left(-\lambda_2 + \left(1+\frac{2}{\nu+1}\right)\lambda_1
-\frac 23 \nu\right).
\end{eqnarray*}
Altogether this means that
\begin{equation}\label{EqTZ}
\begin{array}{lcl}
T'(\theta,y)|_{T=0} &=& Z_y(\theta) \qquad \qquad \textmd{for }\theta \in (0,\tmax) \textmd{ and } y\in\R,\vspace{0.3cm}\cr%
T'(0,1) &=& (\nu+1)^{-1} Z_1(0).%
\end{array}
\end{equation}
The analysis of $Z_y$ is somewhat lengthy and will be postponed to Section \ref{SectionAnalysisOfZ}. The information
that we need about $Z_y$ is:
\begin{Lemma}[Properties of $Z_y$] \label{LemmaPropsOfZ}
$\,$
\begin{itemize}
\item[a)] There is no pair $\theta,y$ with $\theta \in (0,\tmax)$ and $0<y<1$ such that $Z'_y(\theta) = 0$ and $Z_y''(\theta)
\le 0$.
\item[b)] The function $Z_1(\theta)$ is strictly increasing in $\theta$ on the interval $(0,\tmax)$.
\end{itemize}
\end{Lemma}
Lemma \ref{LemmaPropsOfZ} will be proven in Section \ref{SectionAnalysisOfZ}. Now we have all information at hand to
prove the equations (\ref{Eqq1}), (\ref{Eqq2}) and (\ref{Eqq3}).

\begin{Fact} \label{Factq0}
The function $q(\theta)$ is non-negative on the interval $[0,\tmax]$.
\end{Fact}
\begin{proof}
To prove (\ref{Eqq1}) we assume the contrary, i.e., that $q$ drops below zero somewhere on $[0,\tmax]$. Then there are
two points $0<\theta_1 < \theta_2 \le \tmax$ with $q(\theta_1) = q(\theta_2) = 0$ and $q'(\theta_1) \le 0 \le
q'(\theta_2)$.

Suppose first that $\theta_2 < \tmax$. By (\ref{EqRiccatiq}) we have
\begin{equation*}
q'|_{q=0} = \frac{\nu\theta}{\sinh^2\theta} - \epsilon \theta \quad \textmd{for } \theta < \tmax.
\end{equation*}
This is a strictly decreasing function in $\theta$ and therefore a contradiction to $q'(\theta_1) \le q'(\theta_2)$.

Second, we assume $\theta_2 = \tmax$. Again from (\ref{EqRiccatiq}) we get
$$3 \lim_{\theta\rightarrow \tmax} q'(\theta) = \frac{\nu\tmax}{\sinh^2\tmax} - \epsilon \tmax.$$
We thus get the chain of (in)equalities
$$0 \ge q'(\theta_1) = \frac{\nu\theta_1}{\sinh^2\theta_1} - \epsilon \theta_1 > \frac{\nu\tmax}{\sinh^2\tmax} - \epsilon
\tmax = 3q'(\tmax) \ge 0,$$ %
again a contradiction.
\end{proof}

\begin{Fact} \label{Factq1}
For every small enough $\theta > 0$ holds $q(\theta) < 1$.
\end{Fact}
\begin{proof}
We recall that $q(0) = 1$ and $T(0,1) = 0$. Because $Z_1$ is strictly increasing by Lemma \ref{LemmaPropsOfZ} and
because $T'(\theta,1)|_{T=0} = Z_1(\theta)$, it is impossible that $T(\theta,1)$ is equal to zero on a finite interval.
Thus we can limit our consideration to two separate cases:

First, assume $T(\theta,1) < 0$ for all $\theta\in (0,\theta_1)$, where $\theta_1$ is some small positive number.
Differentiating (\ref{EqDefT}) yields
$$\left.\frac{\partial T}{\partial y}\right|_{y=1} = -\frac 1\theta - \nu\coth\theta - 2p(\theta).$$
Because $p(\theta)\rightarrow 0$ for $\theta \rightarrow 0$, there is some $\theta_2>0$ such that
$$\left.\frac{\partial T}{\partial y}\right|_{y=1} < 0 \quad \textmd{for} \quad \theta \in (0,\theta_2).$$
In the definition (\ref{EqDefT}) of $T$ we see that $T(\theta,y)$, viewed as a function of $y$ for some fixed $\theta$,
is just a parabola opening downwards. We conclude that $T(\theta,y)$ is strictly negative on the set
$(0,\min(\theta_1,\theta_2))\times [1,\infty)$ in the $(\theta,y)$-plane. Keeping in mind (\ref{EqqT}), this means that
$q(\theta)$ must be smaller than one for small enough $\theta > 0$.

Second, assume $T(\theta,1) > 0$ for small enough $\theta$. Then $T'(0,1) \ge 0$, which means in view of (\ref{EqTZ})
that $Z_1(0) \ge 0$. By part b) of Lemma\,\ref{LemmaPropsOfZ}, the function $Z_1$ is strictly increasing, such that
$Z_1(\theta) > 0$ for all $\theta \in (0,\tmax)$. We conclude that $T(\theta,1)$ is strictly positive for all $\theta
\in (0,\tmax)$, because otherwise we would have $Z_1(\theta) = T'(\theta,1) \le 0$ at the point where $T(\theta,1)$
crosses zero. Now assume that Fact \ref{Factq1} is not true, i.e., $q(\theta) \ge 1$ for some $\theta
> 0$. Then, to meet the condition $q(\tmax) = 0$, there must be some $\theta_1 \in (0,\tmax)$ with $q(\theta_1) = 1$
and $q'(\theta_1) \le 0$. But this would mean that $T(\theta_1,1) \le 0$, a contradiction to what was said above.
\end{proof}

\begin{Fact} \label{Factq2}
If $0 < q(\theta) < 1$ for some $\theta \in (0,\tmax)$, then $q'(\theta) \le 0.$
\end{Fact}
\begin{proof}
Assume the contrary. Then, because of the boundary behavior (\ref{EqBoundaryq}) of $q$, there must be three points $0 <
\theta_1 < \theta_2 < \theta_3 < \tmax$ such that $0 <  q(\theta_1) = q(\theta_2) =q(\theta_3) =: \tilde q < 1$ and
$q'(\theta_1) < 0$, $q'(\theta_2) > 0$ and $q'(\theta_3) < 0$. For the function $T(\theta,y)$ this means that
$$T(\theta_1,\tilde q) < 0, \quad T(\theta_2,\tilde q) > 0, \quad T(\theta_4,\tilde q) < 0.$$
We also have the boundary behavior
\begin{eqnarray*}
\lim_{\theta\rightarrow 0} T(\theta,\tilde q) &=& +\infty,\\
\lim_{\theta\rightarrow \tmax} T(\theta,\tilde q) &=& +\infty.
\end{eqnarray*}
We conclude that $T(\theta,\tilde q)$ changes its sign at least four times on the interval $(0,\tmax)$. Consequently,
there are four points $\tilde\theta_1 < \tilde\theta_2 <\tilde\theta_3 < \tilde\theta_4$ with
$$Z_{\tilde q} (\tilde \theta_1) \le 0, \quad Z_{\tilde q} (\tilde \theta_2) \ge 0, \quad Z_{\tilde q} (\tilde \theta_3) \le 0,
\quad Z_{\tilde q} (\tilde \theta_3) \ge 0.$$%
Then between $\tilde\theta_1$ and $\tilde\theta_3$ there must be some point $\theta$ such that $q'(\theta) = 0$ and
$q''(\theta) \le 0$, a contradiction to part a) of Lemma \ref{LemmaPropsOfZ}.
\end{proof}

From Fact \ref{Factq0}, Fact \ref{Factq1} and Fact \ref{Factq2} we conclude that the equations (\ref{Eqq2}) and
(\ref{Eqq3}) are true, which proves Lemma \ref{LemmaMonotonicity}.

\section{Analysis of the function $Z_y$} \label{SectionAnalysisOfZ}

This section will be devoted to the proof of Lemma \ref{LemmaPropsOfZ}. Recall that $Z_y$ was defined to be the right
hand side of (\ref{EqTprimeatTzero}). It is straight-forward to simplify this expression to
\begin{equation}\label{EqDefZ}
Z_y = \sum\limits_{i=1}^6 c_i A_i + c_7
\end{equation}
with
\begin{equation*}
\begin{array}{rclrcl}
A_1 &=& -\coth^2\theta, \quad \quad   & A_2 &=& -\theta^{-2}, \vspace{0.1cm} \cr %
A_3 &=& \theta^2\sinh^{-4}\theta,                          & A_4 &=& -\theta^2\sinh^{-2}\theta,  \vspace{0.1cm}\cr %
A_5 &=& -(\theta\coth\theta-1)\sinh^{-2}\theta,  & A_6 &=& \theta^2
\end{array}
\end{equation*}
and
\begin{equation*}
\begin{array}{rclrcl}
c_1 &=& \frac 12\nu^2y, \quad \quad \quad   & c_2 &=& \frac 12(y-y^3), \vspace{0.1cm} \cr %
c_3 &=& \frac 12\nu^2y^{-1},                                & c_4 &=& \epsilon\nu y^{-1},  \vspace{0.1cm}\cr %
c_5 &=& 2\nu,                                    & c_6 &=& \frac 12 \epsilon^2 y^{-1}.
\end{array}
\end{equation*}
The constant $c_7$ includes all terms that do not depend on $\theta$. Thus none of the $c_i$ depends on $\theta$ and
$$c_i > 0 \textmd{ for } i = 1,...,6 \textmd{ and } 0<y<1.$$
\begin{proof}[Proof of Lemma \ref{LemmaPropsOfZ} part a)]
To prove part a) of Lemma \ref{LemmaPropsOfZ} we will show that for $0<y<1$ there is no $\theta>0$ such that
$Z'_y(\theta) \le 0$ and $Z''_y(\theta) \le 0$. We will think of $\theta$ as some fixed number and thus drop it for the
sake of brevity wherever it appears as an argument of the functions $Z_y(\theta)$ or $A_i(\theta)$. Further, we
consider $\vec Z = (Z_y',Z_y'')$ and $\vec A_i = (A_i',A_i'')$ as vectors in the plane. Then we have to show that $\vec
Z$ does not lie in the third quadrant. It is clear that $\vec Z$ is a linear combination with positive coefficients of
the $\vec A_i$. The vector $\vec A_1$ lies in the fourth quadrant because $A_1' > 0$ and $A_1'' < 0$. Let $H$ be the
half-plane that is limited by $\{c \cdot \vec A_1, c \in \mathbb R\}$ and contains the first quadrant. By the Lemma
\ref{LemmaCrossproducts} below we see that the vectors $\vec A_i, i=2,...,6$ lie in the interior of $H$. We conclude
that no linear combinations of the $\vec A_i $ with positive coefficients can lie in the third quadrant, which proves
Lemma \ref{LemmaPropsOfZ}.
\end{proof}

\begin{Lemma}\label{LemmaCrossproducts}
All cross products
$$\vec A_1 \times \vec A_i = A_1'A_i''-A_1''A_i' \quad \quad (i = 2,...,6)$$
are strictly positive.
\end{Lemma}
\begin{proof}[Proof of Lemma \ref{LemmaCrossproducts}]
\noindent For $i=2$ we calculate
$$\vec A_1\times\vec A_2 = \frac{4}{\theta^4\sinh^4 \theta} f(\theta)$$
with
\begin{eqnarray*}
f(\theta) &=& 3\theta \cosh^2\theta - 3 \cosh\theta\sinh\theta-\theta\sinh^2\theta\\
&=& \theta \cosh 2\theta + 2\theta - \frac 32 \sinh 2\theta.
\end{eqnarray*}
The positivity $f$ (and therefore of $\vec A_1\times\vec A_2$) for $\theta > 0$ now follows from $f(0) = f'(0) = 0$ and
$$f''(\theta) = 2\cosh 2\theta (2\theta - \tanh 2\theta) > 0.$$

\noindent For $i=3$ we have
$$
\vec A_1\times\vec A_3 = \frac{4 f(\theta)}{\sinh^9\theta}
$$
with
\begin{eqnarray*}
f(\theta) &=&-6\theta\cosh^2\theta\sinh\theta + \theta\sinh\theta + (4\theta^2+1)\cosh^3\theta -\cosh\theta.
\end{eqnarray*}
It is $f(0)= 0$ and $f'(\theta) = \cosh(\theta) g(\theta)$ with
$$g(\theta) = -3\cosh\theta\sinh\theta-12 \theta\sinh^2\theta+2\theta\cosh^2\theta+\theta+12\theta^2\cosh\theta\sinh\theta
.$$ %
One can check that $g(0) = 0$ and
\begin{eqnarray*}
g'(\theta) &=& 4\sinh\theta (\theta\cosh\theta-\sinh\theta) + 12(\theta^2\cosh^2\theta-\sinh^2\theta)\\
&&+12\theta^2 \sinh^2\theta\\
&>&0
\end{eqnarray*}
Consequently, $g(\theta) \ge 0$ for $\theta \ge 0$ and therefore $f'(\theta) > 0$. It follows that $f(\theta)> 0$ for
$\theta>0$ and finally $\vec A_1\times\vec A_3 > 0$.

\noindent For $i=4$ we have
\begin{eqnarray*}
\vec A_1\times\vec A_4 &=& \frac{4}{\sinh^6\theta}(2\theta\cosh^2\theta-\cosh\theta\sinh\theta-\theta)\\
&=& \frac{4}{\sinh^6\theta}(\theta(\cosh^2\theta-1) + \cosh\theta(\theta\cosh\theta-\sinh\theta))\\
&>& 0.
\end{eqnarray*}

\noindent For $i=5$ we have
$$\vec A_1\times\vec A_5 = \frac{2}{\sinh^8\theta} f(\theta)$$ %
with
\begin{eqnarray*}
f(\theta) &=& -4\theta\cosh^2\theta+2\sinh\theta\cosh^3\theta+\sinh\theta\cosh\theta+\theta\\
&=& -\theta - 2\theta \cosh 2\theta + \sinh 2\theta + \frac 14 \sinh 4\theta
\end{eqnarray*}
We have $f(0) = f'(0) = 0$ and
$$f''(\theta) = 4\sinh 2\theta(\cosh 2\theta-1) + 4\cosh 2\theta (\sinh 2\theta - 2\theta) > 0$$
for $\theta > 0$. We conclude that $f(\theta) > 0$ for $\theta > 0$ and therefore $\vec A_1\times\vec A_5>0$.

\noindent The statement for $i=6$ follows from the fact that $\vec A_6$ lies in the first quadrant ($A_6'
> 0, \, A_6''
> 0$) and $\vec A_1$ lies in the fourth quadrant.
\end{proof}


\begin{proof}[Proof of Lemma \ref{LemmaPropsOfZ} part b)]
We need to show that the function $Z_1(\theta)$ as given by (\ref{EqDefZ}) is strictly increasing in $\theta$. We start
with the partial derivative of $Z_1(\theta)$ by $\epsilon$,
\begin{equation}\label{EqZbyepsilon}
\frac{\partial Z_1}{\partial \epsilon} = \epsilon \theta^2 -\frac{\nu\theta^2}{\sinh^2\theta} + \frac{\partial
c_7}{\partial\epsilon},
\end{equation}
We note that the right hand side of (\ref{EqZbyepsilon}) is an increasing function in $\theta$ and conclude, that if
$Z_1(\theta)$ is increasing for $\epsilon = 0$, then it will be increasing for any $\epsilon > 0$. So it only remains
to show that
\begin{equation}
Z_1(\theta)|_{\epsilon = 0} =2\nu \frac{1-\theta\coth\theta}{\sinh^2\theta} + \frac{\nu^2}{2} \left(\frac{\theta^2}{
\sinh^{4}\theta} - \coth^2\theta \right) + c_7
\end{equation}
is increasing in $\theta$. We set
$$f_1(\theta) = \frac{1-\theta\coth\theta}{\sinh^2\theta}$$
and
$$f_2(\theta) = \frac{\theta^2}{\sinh^{4}\theta} - \coth^2\theta$$
and show that $f_1$ and $f_2$ are increasing separately.

First,
\begin{eqnarray*}
f_1'(\theta) &=& \sinh^{-4}\theta (-\frac 32 \sinh 2\theta+2\theta+\theta\cosh 2\theta)\\
&=& \sinh^{-4}\theta \, \left(2\theta + \sum_{j=1,3,5,\dots} \frac{-3/2}{j!} (2\theta)^j + \frac{j/2}{j!}
(2\theta)^j\right)
\end{eqnarray*}
In the sum over $j$ each single term is non-negative for $j \ge 3$. For $j=1$ the terms in the sum cancel the
$2\theta$. We thus have $f_1'(\theta) > 0$ for $\theta > 0$, such that $f_1$ is strictly increasing.

Second,
\begin{eqnarray*}
f_2'(\theta) &=& \frac{2\cosh \theta}{\sinh^5 \theta} (\theta \tanh \theta + \sinh^2 \theta - 2\theta^2)\\
&\ge&  \frac{2\cosh \theta}{\sinh^5 \theta} (\theta (\theta - \theta^3/3) + (\theta+\theta^3/6)^2 - 2\theta^2)\\
&>& 0
\end{eqnarray*}
for $\theta>0$, which means that $f_2$ also is a strictly increasing function.

\end{proof}

\section{Chiti's comparison argument in $\Hn$} \label{SectionChiti}

Lemma \ref{LemmaChiti} which we will state in this section gives the justification for the last step in each of the
chains of inequalities (\ref{EqChain1}) and (\ref{EqChain2}).

Let $\Omega^\star$ be the spherical rearrangement of $\Omega$, i.e., the geodesic ball (centered at the origin of our
coordinate system) with the same $\Hn$-volume as $\Omega$. For any function $f : \Omega \rightarrow \R^+$ we define the
decreasing rearrangement $f^\sharp(s)$ and the spherical decreasing rearrangement $f^\star(\theta,\vec
\chi)=f^\star(\theta)$. The former is a decreasing function from $[0,|\Omega|]$ to $\R^+$ and is equimeasurable with
$f$ in the measure of $\Hn$. The function $f^\star$ is defined on $\Omega^\star$, spherically symmetric, equimeasurable
with $f$ and decreasing in $\theta$. The functions $f^\sharp$ and $f^\star$ are tightly connected by
\begin{equation}
f^\star(\theta,\vec\chi) = f^\sharp(A(\theta)),
\end{equation}
where
\begin{equation} \label{EqDefA}
A(\theta) = nC_n \int_0^\theta \sinh^{n-1} \theta' \di \theta'
\end{equation}
is the volume of a geodesic ball in $\Hn$ with the radius $\theta$. Here $nC_n$ is the surface area of the
$(n-1)$-dimensional unit sphere in Euclidean space. In a completely analogous way we also define the increasing
rearrangements $f_\sharp$ and $f_\star$.

\begin{Lemma}[Chiti comparison result] \label{LemmaChiti}
Let $u_1(\theta,\vec \chi)$ be the first Dirichlet eigenfunction of $-\Delta$ on $\Omega$ and $z_0(\theta)$ the first
Dirichlet eigenfunction of $-\Delta$ on $S_1$, normalized such that
\begin{equation} \label{EqNormal}
\int_\Omega u_1^2 \di V = \int_{S_1} z_0^2 \di V.
\end{equation}
Then there is some $\theta_0 \in (0,\tmax)$ such that
\begin{equation*}
\begin{array}{ll}
z_0(\theta) \ge u_1^\star(\theta) \quad \quad &\textmd{for } \theta\in (0,\theta_0) \textmd{ and}\cr%
z_0(\theta) \le u_i^\star(\theta) & \textmd{for } \theta\in (\theta_0,\tmax).
\end{array}
\end{equation*}
\end{Lemma}
\begin{proof}
Define $\Omega_t = \{x \in \Omega| u_1(x) > t\}$ and $\partial\Omega_t = \{x \in \Omega| u_1(x) = t\}$. Let
$\mu(t)=|\Omega_t|$ and $|\partial\Omega_t| = H_{n-1}(\partial\Omega_t)$, where $H_{n-1}$ denotes the
$(n-1)$-dimensional measure on $\Hn$. Then the co-area formula
\begin{equation} \label{EqO1}
-\mu'(t) = \int_{\partial \Omega_t} \frac{1}{|\Grad u_1|}\di H_{n-1}
\end{equation}
holds (see, e.g., \cite{C}, p. 86). On the other hand, applying Gauss' Theorem (in its form for Riemannian manifolds,
see, e.g., \cite{C}, p. 7) to $-\Delta u_1 = \lambda_1 u_1$ yields
\begin{equation} \label{EqO2}
\int_{\Omega_t} \lambda_1 u_1 \di V = \int_{\partial\Omega_t} |\Grad u_1| \di H_{n-1},
\end{equation}
since the outward normal to $\Omega_t$ is $-\Grad u_1/|\Grad u_1|$. Using the Cauchy-Schwarz inequality and equations
(\ref{EqO1}) and (\ref{EqO2}) we get
\begin{equation} \label{EqO3}
\bigl(H_{n-1}(\partial\Omega_t)\bigr)^2 = \left(\int_{\partial\Omega_t} \di H_{n-1} \right)^2 \le -\mu'(t) \lambda_1
\int_{\Omega_t} u_1 \di V.
\end{equation}
In $\Hn$ the classical isoperimetric inequality holds true \cite{C,Os78}, i.e.,
\begin{equation} \label{EqCII}
H_{n-1}(\partial\Omega_t) \ge H_{n-1}(\partial (\Omega_t^\star)).
\end{equation}
Recall the definition (\ref{EqDefA}) and write $A^{-1}$ for the inverse function of $A$. Then the $(n-1)$-dimensional
measure of $\partial(\Omega_t^\star)$ can in turn be expressed by
\begin{equation}
H_{n-1}(\partial(\Omega_t^\star)) = nC_n \sinh^{n-1}\theta A^{-1}(|\Omega_t^\star|) = A'(A^{-1}(|\Omega_t^\star|)).
\end{equation}
Hence, equation (\ref{EqCII}) turns into
\begin{equation}
H_{n-1}(\partial\Omega_t) \ge A'(A^{-1}(|\Omega_t^\star|))
\end{equation}
and (\ref{EqO3}) can be written as
\begin{equation} \label{EqO4}
\lambda_1 \int_{\Omega_t} u_1 \di V \ge -\frac{1}{\mu'(t)} A'(A^{-1}(|\Omega_t^\star|))^2.
\end{equation}
Finally we use the fact that
\begin{equation} \label{EqO5}
\int_{\Omega_t} u_1 \di V = \int_0^{\mu(t)} u_1^\sharp(s) \di s,
\end{equation}
which follows directly from the definition of $u_1^\sharp$ above. Since $u_1^\sharp(s)$ is the inverse function to
$\mu(t)$, we have
$$-\frac{\di u_1^\sharp}{\di s}= - \frac{1}{\mu'(t)},$$
which combined with (\ref{EqO4}) and (\ref{EqO5}) yields
\begin{equation} \label{EqO6}
-\frac{\di u_1^\sharp}{\di s} \le \lambda_1 A'(A^{-1}(s))^{-2} \int_0^s u_1^\sharp (s') \di s'.
\end{equation}
One can check that equality holds in all the steps leading to (\ref{EqO6}), if one replaces $\Omega$ by $S_1$ and $u_1$
by $z_0$. Therefore,
\begin{equation} \label{EqO7}
-\frac{\di z_0^\sharp}{\di s} = \lambda_1 A'(A^{-1}(s))^{-2} \int_0^s z_0^\sharp (s') \di s'.
\end{equation}
Using the relations (\ref{EqO6}) and (\ref{EqO7}) and keeping in mind the normalization (\ref{EqNormal}), we will prove
that the functions $u_1^\sharp$ and $z_0^\sharp$ are either identical or they cross each other exactly once on the
interval $[0,|S_1|]$. The arguments we use depend on the fact that $u_1^\sharp$ and $z_0^\sharp$ are continuous
functions. By definition of the rearrangement, both functions are decreasing and we have $z_0^\sharp(|S_1|) =
u_1^\sharp(|\Omega|) = 0$. Recall that from the Rayleigh-Faber-Krahn inequality and from $\lambda_1(S_1) =
\lambda_1(\Omega)$ follows that $|S_1| \le |\Omega|$. From the normalization (\ref{EqNormal}) it is clear that
$z_0^\sharp$ and $u_1^\sharp$ are either identical or cross at least once on $[0,|S_1|]$. To show that they cross
exactly once, we assume the contrary, i.e., that they cross at least twice. Then there are two points $ 0 \le s_1 < s_2
< |S_1|$ such that $u_1^\sharp(s) > z_0^\sharp(s)$ for $s \in (s_1,s_2)$, $u_1^\sharp(s_2) = z_0^\sharp(s_2)$ and
either $u_1^\sharp(s_1) = z_0^\sharp(s_1)$ or $s_1=0$. We set
\begin{equation}
v(s) = \left\{\begin{array}{ll}%
u_1^\sharp(s) & \textmd{on } [0,s_1] \textmd{ if } \int_0^{s_1} u_1^\sharp(s) \di s > \int_0^{s_1} z_0^\sharp(s) \di
s,\cr%
z_0^\sharp(s) & \textmd{on } [0,s_1] \textmd{ if } \int_0^{s_1} u_1^\sharp(s) \di s \le \int_0^{s_1} z_0^\sharp(s) \di
s,\cr%
u_1^\sharp(s) & \textmd{on } [s_1,s_2],\cr%
z_1^\sharp(s) & \textmd{on } [s_2,|S_1|].%
\end{array}\right.
\end{equation}
Then one can convince oneself that because of (\ref{EqO6}) and (\ref{EqO7})
\begin{equation} \label{EqO8}
-\frac{\di v}{\di s} \le \lambda_1 A'(A^{-1}(s))^{-2} \int_0^s v(s') \di s'
\end{equation}
for all $s\in[0,|S_1|]$. Now define the test function $\Psi(\theta,\vec\chi) = v(A(\theta))$. Using the Rayleigh-Ritz
characterization of $\lambda_1$, then (\ref{EqO8}) and finally an integration by parts, we get (if $z_0$ and $u_1$ are
not identical)
\begin{eqnarray*}
\lambda_1 \int_{S_1} \Psi^2 \di V &<& \int_{S_1} |\Grad \Psi|^2 \di V\\
&=& \int_0^\tmax \bigl( A'(\theta) \, v'(A(\theta))\bigr)^2 \,A'(\theta) \di \theta\\
&\le& -\int_0^\tmax A'(\theta)\, v'(A(\theta)) \lambda_1 \int_0^{A(\theta)} v(s')
\di s' \di\theta\\
&=& \lambda_1 \int_0^{|S_1|}  v(s)^2 \di s\\
&=& \lambda_1 \int_{S_1} \Psi^2 \di V
\end{eqnarray*}
Comparing the first and the last term in the above chain of (in)equalities reveals a contradiction to our assumption of
two intersections of $u_1^\sharp$ and $z_0^\sharp$, thus proving Lemma \ref{LemmaChiti}.
\end{proof}


\end{document}